\documentstyle [preprint,aps]{revtex}
\input{epsf}
\begin{document}
\baselineskip14pt
\draft
\bibliographystyle{unsrt}

\title{{Random walk on disordered networks}}
\author{{Tomaso Aste} }

\date{\today}
\maketitle

\centerline{\it Equipe de Physique Statistique, LUDFC  Universit\'e Louis Pasteur,}
\centerline{\it 3, rue de l'Univresit\'e, Strasbourg France}
\centerline{\it e-mail: tomaso@fresnel.u-strasbg.fr}
\centerline{PACS: 66.30.-h, 05.40.-j, 61.43.+j}

\begin{abstract}
Random walks are studied on disordered cellular networks in 2-and
3-dimensional spaces with arbitrary curvature.
The coefficients of the evolution equation are calculated
in term of the structural properties of the cellular system.
The effects of disorder and space-curvature on the diffusion
phenomena are investigated.
In disordered systems the mean square displacement displays an enhancement
at short time and a lowering at long ones, with respect to the ordered case.
The asymptotic expression for the diffusion equation on hyperbolic
cellular systems relates random walk on curved lattices to
hyperbolic Brownian motion.
\end{abstract}

\vskip.7cm
\noindent
The simplest disordered networks are space-filling random partitions of space
by cells.
The cells are convex, irregular polygons in two dimension ($2d$)
and irregular polyhedras in $3d$.
Disorder imposes the incidence numbers at their minimum values
($d+1$ edges incident on a vertex in $d$-dimensions).
These cellular networks are also known in the literature as ``froths'', 
since the soap-froth is the archetype of such structures.
The space tiled by the froth can be curved.
In this case, the intrinsic dimension of the cellular system
($d_f$) does not coincide with the dimension ($d$) of the embedding
space.

Froths are the structures which characterizes a broad class of
natural systems such as polycrystalline solids, foams, biological
tissues etc. \cite{WeRi84,Stav93}.
Moreover, froths and disordered packings are dual systems (see
Fig.1).
Therefore, amorphous materials, granular solids, metallic glasses
etc. have structures which are dual of froths
\cite{SadMos85,RivSad87}.

Many theoretical works, experiments, and computer simulations have
been devoted to the study of random walk and transport phenomena
on disordered \cite{Lim90,Zan95,Dur94Vera96,McMurry94} and fractal
\cite{Bun91,Comt96,Boe95,Bend95,Cassi96,Bress96,Mont96} systems.
Random walk on Euclidean froths is a realistic model for diffusion
in disordered systems, for signal propagation in granular media
and it can be relevant in the study of the evolution properties of
natural foams and polycrystalline aggregates.
Whereas, random walk on hyperbolic or elliptic froths can model
transport phenomena in curved spaces.

Disordered structures are widespread in nature, and natural,
disordered, two dimensional cellular structures can tile curved
surfaces (e.g. amphiphilic membranes or epithelial tissues).
The study of the effect of disorder and space-curvature on the
diffusion phenomena is therefore of great physical interest and it
is the aim of this paper.

\bigskip
In the present model, the walker starts at time $t=0$ from a given
cell, then, at each (finite) time-step it jumps with 
equal-probability to one of the neighbouring cells.
The radial and angular components of the motion respect with the
starting cell are decoupled.
The radial component results the same as of the
spherically symmetric model introduced recently in
\cite{Boe95,Bend95}.
But, in the present case, the diffusion is on realistic cellular
systems and all the parameters in the evolution equation are given
in terms of the properties of the disordered cellular structure.

A $d$-dimensional disordered cellular froth, can be studied as structured in
concentric layers of cells at the same topological distance ($j$)
around a given central cell (where the topological distance
between two cells is the minimum number of $(d-1)$-dimensional
interfaces that a path must cross to connect the two cells).
The structure is described topologically, by two parameters per layer in
$2d$ (number of cells per layer and average coordination in the layer), 
and three parameter per layer in $3d$ (see \cite{AsBoRi} for details).

The number of cells in a layer at distance $j$ from the central
cell ($K(j)$), is related to the space-curvature.
One finds asymptotically, $K(j) \propto j^{d_f-1}$, where $d_f$ is
the intrinsic dimension of the cellular system.
The intrinsic dimension $d_f$ coincide with the dimension $d$ of
the embedding space in Euclidean froths (which are cellular
tilings of flat spaces), whereas, $d_f > d$ in the hyperbolic case
(tilings of negatively curves spaces) and $d_f < d$ in the
elliptic one (tilings of positively curves spaces).
A special case, discussed in \cite{AsBoRi}, is a class of
hyperbolic froths with $K(j) \propto \exp(\varphi j)$.
Here the intrinsic dimension diverges.

Suppose the system being shell-structured-inflatable (SSI) around
the central cell.
(In SSI froths any cell in layer $(j)$ has neighbours in layer 
$(j-1)$, $(j)$ and $(j+1)$ and the layers make, concentric, closed
rings of cells without ``topological defects''. See Fig.1 and
\cite{AsBoRi} for details.)
The number of paths connecting different layers can be more easily
calculated in SSI froths.
The extension to the general case of non-SSI froths, follows
straightforward.

Let the central cell be the one where the walker starts at $t=0$.
Consider at time $t$ the walker being in a cell of layer $(j)$ (the
cells in layer $j$ are supposed indistinguishable).
At time $t+1$ it has moved outwards to layer $(j+1)$ or (for $j >0$)
inwards to layer $(j-1)$ or within the same layer $(j)$, with
probabilities $p_{out}(j)$ or $p_{in}(j)$ or $p_{stay}(j)$,
respectively.
Note that, $p_{in}(j)+p_{out}(j)+p_{stay}(j)=1$, since the walker
must move at each time-step.

The probability $P(j,t)$ of finding the walker in layer $(j)$ at
time $t$ is given by
\begin{eqnarray}
P(j,t)&=&p_{stay}(j)P(j,t-1)+p_{out}(j-1)P(j-1,t-1) \nonumber \\
&+& p_{in}(j+1)P(j+1,t-1)
\;\;\;\;\;\;\;\;\;\;\;\;\;\;\;\;\;\;\;\;\;\;\;\;\;\;\;\;\;\;\;\;
\mbox{for $j \ge 1$ },
\label{prob1}
\end{eqnarray}
and, for $j=0$, by
\begin{equation}
P(0,t)=p_{in}(1)P(1,t-1)\;\;\;\;.
\label{prob10}
\end{equation}
With initial conditions $P(j,0)= \delta_{j,0}$.

The probability $p_{out}(j)$ is proportional to the number of
paths connecting layer $(j)$ with layer $(j+1)$.
This number is equal to the number of interfaces (edges in $2d$
and facets in $3d$) separating the two layers.
Analogously, the probability $p_{in}(j)$ is proportional to the
number interfaces between layer $(j)$ and $(j-1)$.

In $2d$, the number of edges separating layers $(j)$ and $(j+1)$
is $K(j)+K(j+1)$ (see fig.1).
Thus,
\begin{equation}
\begin{array}{l}
p_{out}(j) = {1 \over {\cal N}_2(j)}\Big(K(j)+K(j+1) \Big) \\
p_{in}(j) = {1 \over {\cal N}_2(j)}\Big(K(j)+K(j-1) \Big)
\end{array}
  \;\;\;\;\;\;\;\;   \mbox{(for $j \ge 1$)},
\label{2probjump}
\end{equation}
(and $p_{stay}(j)=1-p_{out}(j)-p_{in}(j) = {2 \over e(j)}$).
In Eq.(\ref{2probjump}) we defined, $K(0)=0$ and ${\cal
N}_2(j)=e(j) K(j)$, with $e(j)$ the average number of edges per
cell in layer $(j)$.
For $j=0$, one has $p_{out}(0)=1$ and $p_{in}(0)=0$.

In $3d$ the layers are separated by a system of faces which tile
a spherical surface: the ``shell-network'' \cite{AsBoRi}.
The number of paths between two successive layers $(j)$ and
$(j+1)$ is proportional to the number of facets of the 
shell-network between these layers.
This number is $2 {K(j)+K(j+1) -8 \over n(j) -4 }$ \cite{AsBoRi},
where $n(j)$ is the average number of edges per face in the shell-network.
We have therefore
\begin{equation}
\begin{array}{l}
p_{out}(j) = {2 \over {\cal N}_3(j)}{\displaystyle K(j)+K(j+1) -8
\over n(j) -4 }    \\
p_{in}(j) = {2 \over {\cal N}_3(j)}{\displaystyle K(j)+K(j-1) -8
\over n(j-1) -4 }
\end{array}
\;\;\;\;\;\;\;\;\;\;  \mbox{(for $j \ge 1$)},
\label{3probjump}
\end{equation}
where we defined: $K(0)= 2$ and ${\cal N}_3(j)=f(j) K(j)$, with
$f(j)$ the average number of faces of the cells in layer $(j)$.
For $j=0$, we have $p_{out}(0)=1$ and $p_{in}(0)=0$.

\bigskip
A quantity of interest is the probability $\Pi$ that the walker
ever returns to the origin.
This quantity is associated with the mean time spent in the origin
($F(0)=\sum_{t=0}^{\infty}P(0,t)$) by the relation  $\Pi=1-{1
\over F(0)}$ \cite{ItzSFT}.
From Eq.(\ref{prob1}) and by using Eqs.(\ref{2probjump}) and
(\ref{3probjump}), we obtain
\begin{equation}
\Pi=1 - {\displaystyle 1 \over F(0)}=1-{\displaystyle 1 \over 1 +
K(1) \sum_{j=1}^{\infty} {1 \over {\cal
N}_d(j)p_{out}(j) } }\;\;\;\; .
\label{P}
\end{equation}
This expression is valid for any froth tiling an unbounded topological 
manifold.
The quantity ${\cal N}_d(j)p_{out}(j)$ is related to the
properties of the structure around the central cell, and asymptotically
it scales as $K(j)$ (see Eqs.(\ref{2probjump}) and (\ref{3probjump})).
In a cellular system with intrinsic dimension $d_f$ the number of
cells per layer have the asymptotic behaviour $K(j) \propto j^{d_f-
1}$, thus ${\cal N}_d(j)p_{out}(j) \sim K(j) \sim  j^{d_f-1}$.
By substituting into Eq.(\ref{P}), we obtain $\Pi =1$ for 
$d_f \le 2$, and $\Pi < 1$ for $d_f > 2$.
This result, already known for random walk on regular lattices,
fractals, trees \cite{ItzSFT}, and found here in froths, indicates
the universality of this critical behaviour which is independent
of the details of the structure.
In Fig.2 the behaviour of $\Pi$ vs. $d_f$, given by equation
(\ref{P}) for $2d$ SSI froths with $K(j) = K(1) j^{d_f-1}$, is
shown.

\bigskip

A quantity generally used to describe the diffusion phenomena is
the mean square displacement $\langle r^2
\rangle(t)=\sum_{j=0}^{\infty} j^2 P(j,t)$.
The time-dependent diffusion coefficient $D(t)$ is associated with
this quantity  by the relation $ 2d D(t) = {\partial \over
\partial t} \langle r^2 \rangle  $, and the usual diffusion
coefficient $D^\infty$ is the limit of $D(t)$ at infinite time.

From Eq.(\ref{prob1}) follows
\begin{equation}
\langle r^2 \rangle(t+1) - \langle r^2 \rangle(t) =
\sum_{j=0}^\infty \Bigg\{p_{out}(j) + p_{in}(j) + 
2 j \Big[ p_{out}(j) - p_{in}(j) \Big] \Bigg\} P(j,t)
\;\;\;\;\;.
\label{<j^2>}
\end{equation}
When, $j \gg 1$, and the parameters $e(j)= \langle e \rangle$,
$f(j)= \langle f \rangle$ and $n(j)= \langle n^N \rangle$ are
independent of $j$ (this is the expected asymptotic behaviour), 
Eqs.(\ref{2probjump}) and (\ref{3probjump}) give
\begin{equation}
p_{out}(j)+p_{in}(j)  = 1- p_{stay}(j) = 
\left\{ \begin{array}{c} 
{\langle e \rangle -2 \over  \langle e \rangle}
\;\;\;\; \mbox{for $d=2$} \\
{\langle f \rangle -6 \over  \langle f \rangle} \;\;\;\;
\mbox{for $d=3$}   \end{array}
\right.
= 2{\cal C}_d  \;\;\; ,
\label{p+p}
\end{equation}
and, for  $d_f$ finite, $j [p_{out}(j) - p_{in}(j)]= (d_f-1) {\cal C}_d$. 
Thus, from Eq.(\ref{<j^2>})
\begin{equation}
\langle r^2 \rangle(t) \sim  2 d_f {\cal C}_d t \;\;\;\;\;\;.
\label{D2}
\end{equation}
The diffusion coefficient is therefore, $D^\infty={d_f \over d} {\cal C}_d$. 
Numerical solutions of Eq.(\ref{prob1}) for $2d$ and $3d$
structures with different intrinsic dimensions and coordinations
give diffusion coefficients in very good agreement with Eq.(\ref{D2}).
Note that, $\langle r^2 \rangle$ in Eq.(\ref{D2}) is expressed in term 
of topological distances. 
The metric quantities can be retrieved from the 
topological ones by multiplying the topological distance $j$ by 
the average asymptotic distance $\rho_0$ between layers. 
For instance, in the hexagonal lattice, $\rho_0 = {\sqrt 3 \over 2} a$, 
with $a$ the lattice spacing. 
From Eq.(\ref{D2}), one gets therefore 
$\langle \rho^2 \rangle = \rho_0^2 \langle r^2 \rangle = a^2 t$,
which is  the known expression for the mean square displacement in 
the hexagonal lattice.

The linear dependence on $t$ of $\langle r^2 \rangle$ in Eq.(\ref{D2}),
indicates a diffusive behaviour.
In this case, the spectral dimension $d_s$ (defined from the
exponents $\langle r^2 \rangle \sim t^{d_s /d_f}$ and $P(0,t) \sim
t^{- d_s/2}$ \cite{Alex82}) coincide with the intrinsic dimension
$d_f$.

In disordered froths, topological non-SSI defects are always
present. Defects, in layer $j$, are cells which have no neighbours 
in layer $j+1$ (see fig.1 and \cite{AsBoRi}).
Asymptotically, the number of defective cells in layer $j$ is a fraction 
$\delta$ of the total number of cells $ K(j)$.
Typically, $2d$ Euclidean, disordered froths have $0.1 < \delta < 0.2$ 
 \cite{AST}.
In $2d$, the number of paths connecting layer $j$ with non-defective 
cells in layer $j+1$ is $(1-\delta)(K(j)+K(j+1))$ (see fig.(1)). 
Whereas, the number of paths ending in a defective cell is $\eta \delta K(j+1)$, 
with $\eta$ the average number of interfaces added by a defect to the shell
 between two successive layers (typically, $ 1 < \eta < 1.3$ in $2d$ \cite{AST}).
Therefore, asymptotically, Eqs.(\ref{D2}) and  (\ref{p+p}) can be 
extended to the non-SSI case by multiplying expression (\ref{p+p}) 
by the factor $(1-\delta + {\eta \delta \over 2})$. 
The same result holds in $3d$.

The non-SSI defects have important effects on the froth
structure.
In particular, it has been found in \cite{AST} that, in $2d$ 
non-SSI Euclidean froths, the number of cells per layer increases with
the distance following a linear law $K(j)=Cj+B$, with a slope $C \sim 9$. 
This slope is higher than the both values, $C=2 \pi$ expected
from simple geometrical consideration and $C=6$ of the SSI
hexagonal lattice.
An higher increment in the number of cells per layer,  must
correspond to a {\it faster diffusion} (higher number of paths to go
outward).
On the other hand, in typical $2d$ disordered systems, one has
$(1-\delta + {\eta \delta \over 2})<1$, which indicate asymptotically, a
{\it slower diffusion} in non-SSI froths compared with the ordered 
SSI case.
These two opposite behaviours are not in contradiction.
Indeed, for $j \gg 1$ the ratio between the number of paths in 
successive layers depends only on the
exponent of $K(j)$ vs. $j$ (i.e. the intrinsic dimension - 1), and
not on the slope.
Therefore, we expect the diffusion in disordered structures
compared with ordered SSI lattices, being {\it faster at small distances}
 (where the slope of $K(j)$ is relevant) and then
becoming {\it slower in the asymptotic part} (where only the exponent of
$K(j)$ is relevant).
Fig.3 shows ${\langle r^2 \rangle/ t}$ calculated numerically for
a non-SSI $2d$ Euclidean froth (a) and for the SSI hexagonal
lattice (b).
The diffusion in the disordered structure is faster than in the
SSI hexagonal lattice in the first stage ($t < 15$ and $j < 4$),
then it slows down to reach an asymptotic behaviour
where the mean square displacement grows with $t$ more slowly in
the disordered than in the ordered system.

\bigskip
A special behaviour of $\langle r^2 \rangle$ is obtained for the
$2d$ SSI hyperbolic froth, studied in
Ref.\cite{AsBoRi}, which has $e(j)=\langle e \rangle > 6$ and
$K(j) = C \exp( \varphi j)$, with $\varphi = \cosh^{-1}({\langle e
\rangle -4 \over 2})$.
In this case, from Eqs.(\ref{2probjump}), (\ref{3probjump})
and (\ref{<j^2>}),  one derives the asymptotic expression
\begin{equation}
\langle r^2 \rangle(t) \sim
{(\langle e \rangle -6 )(\langle e \rangle - 2 )\over \langle e
\rangle^2 } t^2
\;\;\;\;\;\mbox{for $t \gg 1$}.
\label{<j^2>(t)}
\end{equation}
Numerical solutions of Eq.(\ref{prob1}) for SSI $2d$ hyperbolic
froths with various $\langle e \rangle >6$, give  time-dependent
diffusion coefficient in excellent agreement with expression
(\ref{<j^2>(t)}).
The quadratic exponent in Eq.(\ref{<j^2>(t)}) indicates a
ballistic diffusion and $d_s = 2 d_f$.

\bigskip

We now write the evolution equation (\ref{prob1}) in the
continuous limit.
Let introduce the continuous variables $\rho = j \rho_0$ and $\tau
= t \tau_0$, where $\rho_0$ is the average distance between two
layers and $\tau_0$ is the average time between two jumps.
In the asymptotic limit ($j = \rho/\rho_0 \rightarrow \infty$ and
$t = \tau/\tau_0 \rightarrow \infty$), when the average
topological arrangements of the cells is independent of the layer
number, equation (\ref{prob1}) can be written in the continuous
form
\begin{equation}
{\displaystyle \partial \over \partial \tau} P(\rho,\tau) =
{\rho_0^2 \over \tau_0} {\cal C}_d {\displaystyle \partial \over
\partial \rho}
\Bigg\{ {\displaystyle \partial \over \partial \rho} P(\rho,\tau)-
\Bigg[ {4 \over (s+2)}{1 \over K(\rho)}
{\displaystyle \partial \over \partial \rho}K(\rho)\Bigg]
P(\rho,\tau)
\Bigg\}\;\;,
\label{Pcont}
\end{equation}
where $s$ is the inflation parameter ($s = \langle e \rangle -4$ in $2d$, and 
$s= {1 \over 2} (\langle f \rangle -6 )(\langle n^N \rangle - 4) -2$ in $3d$ 
\cite{AsBoRi}), which is a
quantity associated with the curvature of the manifold tiled by
the froth ($s=2$ Euclidean, $s>2$ hyperbolic and $s<2$ elliptic froths).
Expression (\ref{Pcont}) is the diffusion equation, for a 
$d$-dimensional spherically symmetric cellular system, written in
polar coordinate.
All the information about the structure, its intrinsic dimension,
the disorder, are contained in the term in the square brackets and
in the parameter ${\cal C}_d$.

For a random cellular system with finite intrinsic dimension $d_f$
we have asymptotically $K(\rho) \sim \rho^{d_f-1}$ and $s
\rightarrow 2$.
Therefore the coefficient inside the square brackets  in
(\ref{Pcont}) becomes ${d_f-1 \over \rho}$, and Eq.(\ref{Pcont})
has the solution
\begin{equation}
P(\rho,\tau) = {\displaystyle 2 \rho^{d_f-1} \over
\Gamma({d_f \over 2}) (4 {\rho_0^2 \over \tau_0} {\cal C}_d
\tau)^{d_f \over 2}}
\exp \Bigg(- {\displaystyle \rho^2 \over 4 {\rho_0^2 \over \tau_0}
{\cal C}_d \tau}
\Bigg)\;\;\;\;.
\label{Psol2}
\end{equation}
The probability $P(\rho,\tau)$ increases with $\rho$ 
until a maximum at $\rho_{max}=(2(d_f -1){\rho_0^2 \over
\tau_0}{\cal C}_d {\tau })^{1 \over 2}$, then it decreases exponentially.
From solution (\ref{Psol2}), the mean square displacement results
$\langle r^2 \rangle \simeq \int_0^\infty {\rho^2 \over \tau_0}
P(\rho,\tau) d \rho = 2 d_f {\rho_0^2 \over \tau_0} {\cal C}_d
\tau$,  as already obtained in Eq.(\ref{D2}).

\bigskip
In a previous paper \cite{AsBoRi}, it was found a class of $2d$
and $3d$ hyperbolic SSI froths which can be generated iteratively
by using a simple recursive equation.
These froths have $K(\rho) = C \sinh (\varphi {\rho \over
\rho_0})$, where $\varphi=\cosh^{-1}({s \over 2})$ is a constant
associated with the space curvature (in simple $2d$ cases one can
show the equivalence $\varphi = \sqrt{ - k}$, with ${ k}$ the
Gaussian curvature. Here, $s > 2$ and $k<0$).
For these froths, the term in the square brackets in
Eq.(\ref{Pcont}) is ${4 \over (s+2)} {\varphi \over \rho_0} \mbox{
coth}( \varphi {\rho \over \rho_0}) $.
Therefore, the evolution equation (\ref{Pcont}) takes the form
\begin{equation}
{\displaystyle \partial \over \partial \tau} P(\rho,\tau) =
{\rho_0^2 \over \tau_0} {\cal C}_d {\displaystyle \partial \over
\partial \rho}
\Bigg\{ {\displaystyle \partial \over \partial \rho} P(\rho,\tau)-
\Bigg[ {4 \over (s+2)} {\varphi \over \rho_0}\mbox{ coth}( \varphi
{\rho \over \rho_0})   \Bigg]
P(\rho,\tau)
\Bigg\}\;\;.
\label{Pcont2}
\end{equation}
Equation (\ref{Pcont2}) was already known in literature as the
diffusion equation in hyperbolic spaces with constant, negative
curvature  \cite{Mont96,Dav}.
Here the same equation have been obtained starting from a
tessellation model, making therefore a link between diffusion in
curved lattices and hyperbolic Brownian motion.
At long distances, the coefficient in the square brackets in
Eq.(\ref{Pcont2}) tends to a constant and the corresponding
solution is a Gaussian with a constant drift.
Here, the probability distribution move ballistically outward with
the maximum at $\rho_{max}={4 \over s+2} \varphi {\rho_0 \over
\tau_0} {\cal C}_d \tau$.
This ballistic diffusion is consistent with the result for the
mean square displacement for hyperbolic froth given in
Eq.(\ref{<j^2>(t)}).

\bigskip
The author acknowledges discussions and correspondence with A.
Comtet, D.J. Durian, E. Galleani d'Agliano, J.F. Joanny, F.
Napoli and N. Rivier.
This work was partially supported by EU, HCM contract
ERBCHRXCT940542 and by TMR contract ERBFMBICT950380.

\newpage
\begin{figure}
\vspace*{-1.cm}
\hspace*{.8cm}
\epsfxsize=12.cm
\epsffile{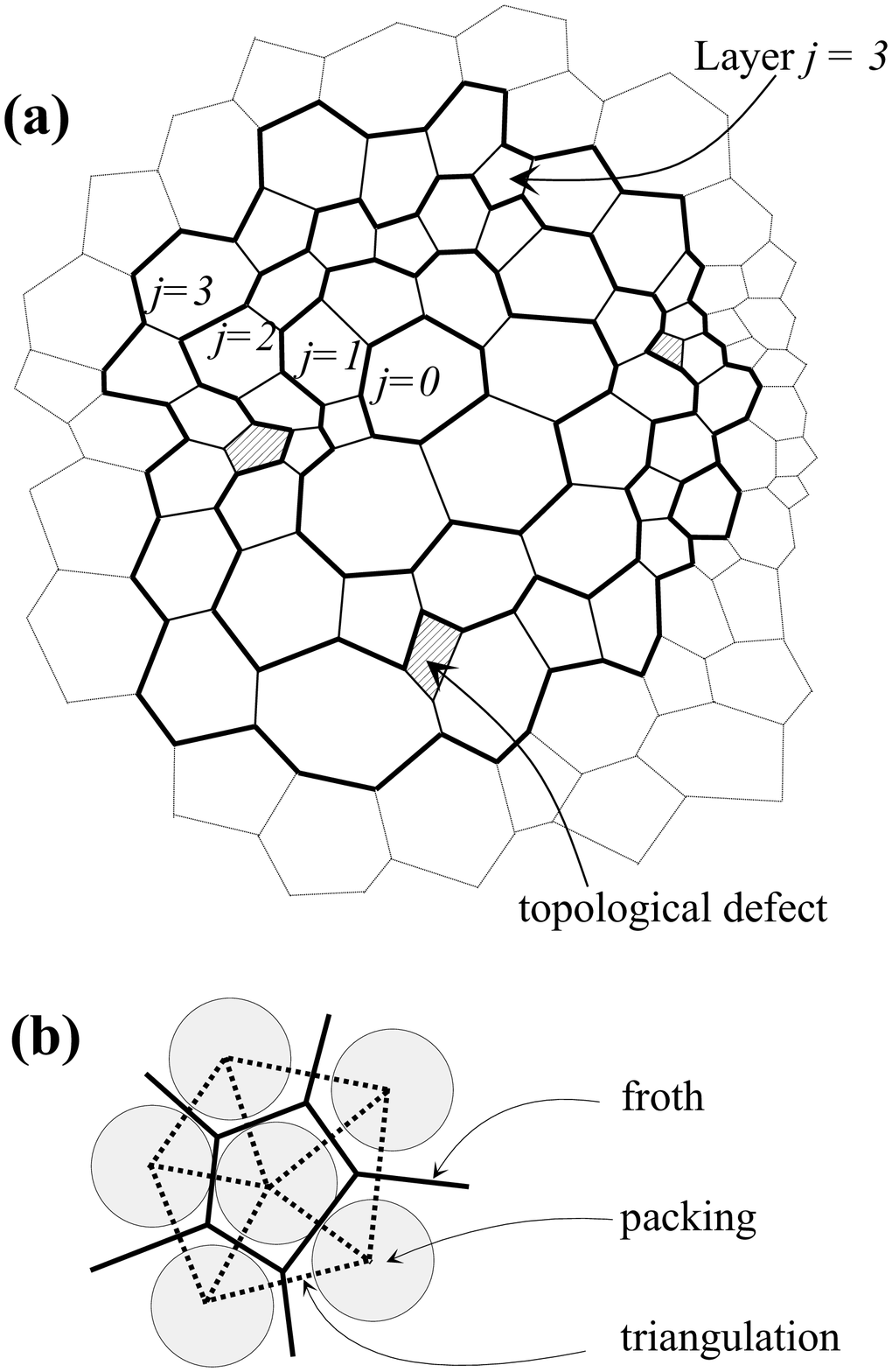}
\label{fig1}
\vspace{-1.cm}
\caption{
A froth is a random partition of space by cells {\bf (a)}.
Disorder imposes the incidence number at the minimum value (3
edges incident on a vertex in $2d$).
Froths are the dual structures of disordered packings {\bf (b)}.
Such structures can be analysed as organized in concentric layers
of cells at the same topological distance ($j$) from a given
central cell ($j=0$).
Some cells (brought out by hatcheries in {\bf (a)}) have
neighbours in the internal layer but not in the external one and
are topological inclusion or ``defects'' in the layered-structure.
}
\end{figure}

\newpage
\begin{figure}
\vspace*{-5.cm}
\epsfxsize=14.cm
\epsffile{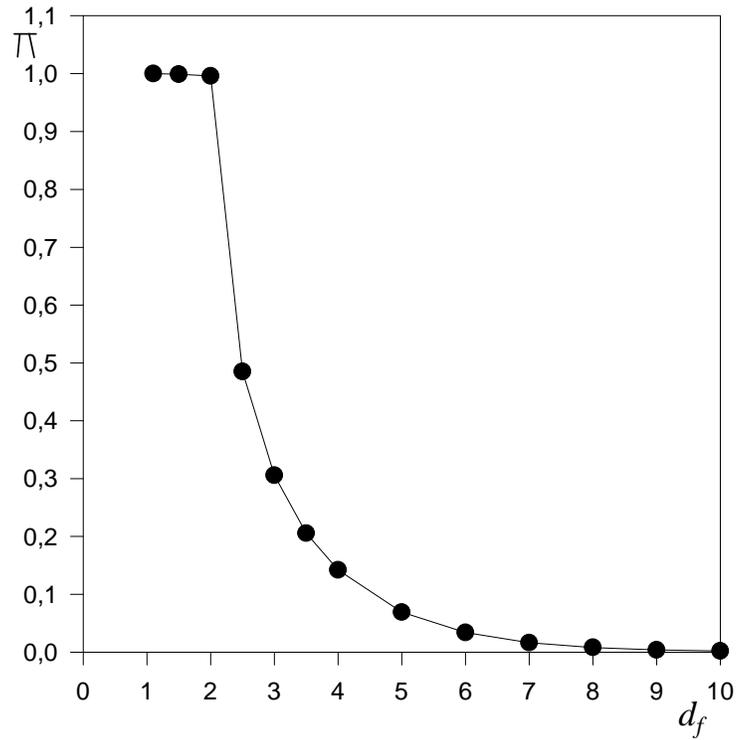}
\label{fig2}
\vspace{-2.cm}
\caption{
Probability $\Pi$ that the walker ever return to the origin for
several values of the intrinsic dimension $d_f$.
The walker always return in the origin when $d_f \le 2$, whereas
the probability  is less than 1 and decreases with $d_f$ when $d_f
> 2$.
This critical behaviour is independent of the details of the
structure.
(The line is a guide for eyes.)
}
\end{figure}

\newpage
\begin{figure}
\vspace*{-5.cm}
\epsfxsize=14.cm
\epsffile{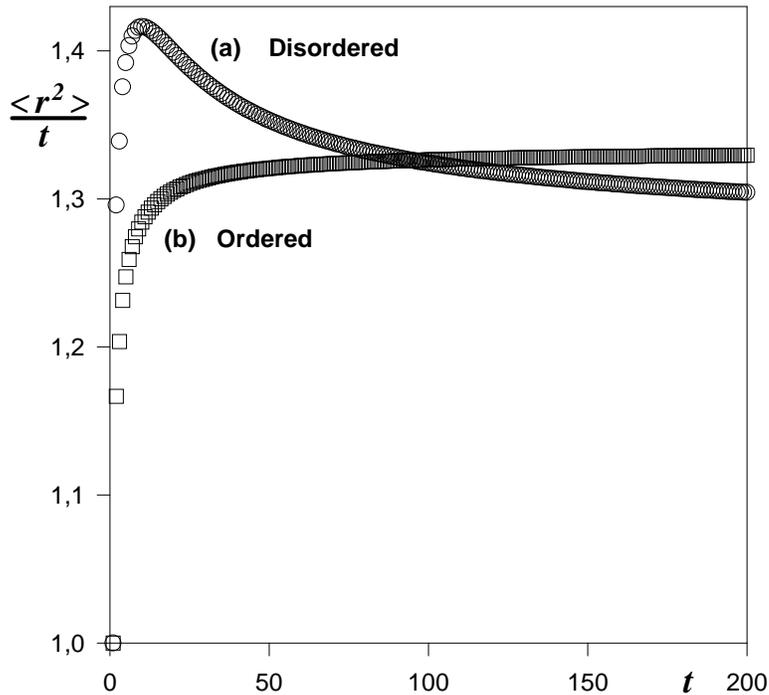}
\label{fig4}
\vspace*{-2.cm}
\caption{
Mean square displacement $\langle r^2 \rangle$ over $t$ vs. time
for disordered {\bf (a)} and ordered {\bf (b)} cellular systems.
The average distance of the walker from the starting point is $j
\simeq \langle r^2 \rangle^{1/2}$.
At short distances ($j < 5$) the walker diffuses faster in
disordered system than in the corresponding ordered lattice.
Then diffusion in disordered system slows down to
reach an asymptotic regime where the walker propagates more slowly in
disordered system than in the ordered case.
}
\end{figure}

\end{document}